\documentclass[12pt,a4paper]{article}
\usepackage{psfrag}
\usepackage{graphics}
\usepackage{amssymb,epsfig,amsmath,euscript,array}
\makeatletter
\@addtoreset{equation}{section}
\makeatother

\newcounter{multieqs}

\newcommand{\be}{\begin{equation}}
\newcommand{\ee}{\end{equation}}

\newcommand{\bm}[1]{\mbox{\boldmath $#1$}}

\def\bd{\begin{document}}
\def\ed{\end{document}}
\def\nn{\nonumber}
\def\bea{\begin{eqnarray}}
\def\eea{\end{eqnarray}}
\let\bm=\bibitem
\let\la=\label

\newcommand{\EQ}[1]{\begin{equation} #1 \end{equation}}
\newcommand{\AL}[1]{\begin{subequations}\begin{align} #1 \end{align}\end{subequations}}
\newcommand{\SP}[1]{\begin{equation}\begin{split} #1 \end{split}\end{equation}}
\newcommand{\ALAT}[2]{\begin{subequations}\begin{alignat}{#1} #2 \end{alignat}\end{subequations}}
\def\beqa{\begin{eqnarray}}
\def\eeqa{\end{eqnarray}}
\def\beq{\begin{equation}}
\def\eeq{\end{equation}}

\def\hf{{\textstyle{1\over2}}}
\def\wbar{\bar w}
\def\mubar{\bar\mu}
\def\abar{\bar a}
\def\sigmabar{\bar\sigma}
\def\etabar{\bar\eta}
\def\zetabar{\bar\zeta}
\def\mubar{\bar\mu}
\def\nubar{\bar\nu}
\def\N{{\cal N}}
\def\sst{\scriptscriptstyle}
\def\thetabar{\bar\theta}
\def\Tr{{\rm Tr}}
\def\one{\mbox{1 \kern-.59em {\rm l}}}
 \def\Nh{\hat{N}}

\def\a{\alpha}      \def\da{{\dot\alpha}}
\def\b{\beta}       \def\db{{\dot\beta}}
\def\c{\gamma}  \def\G{\Gamma}  \def\cdt{\dot\gamma}
\def\d{\delta}  \def\D{\Delta}  \def\ddt{\dot\delta}
\def\e{\epsilon}        \def\vare{\varepsilon}
\def\f{\phi}    \def\F{\Phi}    \def\vvf{\f}
\def\h{\eta}
\def\k{\kappa}
\def\l{\lambda} \def\L{\Lambda}
\def\m{\mu} \def\n{\nu}
\def\o{\omega}
\def\p{\pi} \def\P{\Pi}
\def\r{\rho}
\def\s{\sigma}  \def\S{\Sigma}
\def\t{\tau}
\def\th{\theta} \def\Th{\Theta} \def\vth{\vartheta}
\def\X{\Xeta}
\def\z{\zeta}

\def\cA{{\cal A}} \def\cB{{\cal B}} \def\cC{{\cal C}}
\def\cD{{\cal D}} \def\cE{{\cal E}} \def\cF{{\cal F}}
\def\cG{{\cal G}} \def\cH{{\cal H}} \def\cI{{\cal I}}
\def\cJ{{\cal J}} \def\cK{{\cal K}} \def\cL{{\cal L}}
\def\cM{{\cal M}} \def\cN{{\cal N}} \def\cO{{\cal O}}
\def\cP{{\cal P}} \def\cQ{{\cal Q}} \def\cR{{\cal R}}
\def\cS{{\cal S}} \def\cT{{\cal T}} \def\cU{{\cal U}}
\def\cV{{\cal V}} \def\cW{{\cal W}} \def\cX{{\cal X}}
\def\cY{{\cal Y}} \def\cZ{{\cal Z}}

\def\ua{\underline{\alpha}}
\def\ub{\underline{\phantom{\alpha}}\!\!\!\beta}
\def\uc{\underline{\phantom{\alpha}}\!\!\!\gamma}
\def\um{\underline{\mu}}
\def\ud{\underline\delta}
\def\ue{\underline\epsilon}
\def\una{\underline a}\def\unA{\underline A}
\def\unb{\underline b}\def\unB{\underline B}
\def\unc{\underline c}\def\unC{\underline C}
\def\und{\underline d}\def\unD{\underline D}
\def\une{\underline e}\def\unE{\underline E}
\def\unf{\underline{\phantom{e}}\!\!\!\! f}\def\unF{\underline F}
\def\unm{\underline m}\def\unM{\underline M}
\def\unn{\underline n}\def\unN{\underline N}
\def\unp{\underline{\phantom{a}}\!\!\! p}\def\unP{\underline P}
\def\unq{\underline{\phantom{a}}\!\!\! q}
\def\unQ{\underline{\phantom{A}}\!\!\!\! Q}
\def\unH{\underline{H}}

\def\As {{A \hspace{-6.4pt} \slash}\;}
\def\bs {{b \hspace{-6.4pt} \slash}\;}
\def\Ds {{D \hspace{-6.4pt} \slash}\;}
\def\ds {{\del \hspace{-6.4pt} \slash}\;}
\def\ss {{\s \hspace{-6.4pt} \slash}\;}
\def\ks {{ k \hspace{-6.4pt} \slash}\;}
\def\ps {{p \hspace{-6.4pt} \slash}\;}
\def\pas {{{p_1} \hspace{-6.4pt} \slash}\;}
\def\pbs {{{p_2} \hspace{-6.4pt} \slash}\;}

\def\Fh{\hat{F}}
\def\Vh{\hat{V}}
\def\Xh{\hat{X}}
\def\ah{\hat{a}}
\def\xh{\hat{x}}
\def\yh{\hat{y}}
\def\ph{\hat{p}}
\def\xih{\hat{\xi}}

\def\psit{\tilde{\psi}}
\def\Psit{\tilde{\Psi}}
\def\tht{\tilde{\th}}

\def\At{\tilde{A}}
\def\Qt{\tilde{Q}}
\def\Rt{\tilde{R}}
\def\Nt{\tilde{N}}

\def\at{\tilde{a}}
\def\st{\tilde{s}}
\def\ft{\tilde{f}}
\def\pt{\tilde{p}}
\def\qt{\tilde{q}}
\def\vt{\tilde{v}}
\def\nt{\tilde{n}}

\def\delb{\bar{\partial}}
\def\bz{\bar{z}}
\def\bD{\bar{D}}
\def\bB{\bar{B}}

\def\bk{{\bf k}}
\def\bl{{\bf l}}
\def\bp{{\bf p}}
\def\bq{{\bf q}}
\def\br{{\bf r}}
\def\bx{{\bf x}}
\def\by{{\bf y}}
\def\bR{{\bf R}}
\def\bV{{\bf V}}

\def\d{\delta}\def\D{\Delta}\def\ddt{\dot\delta}

\def\pa{\partial} \def\del{\partial}
\def\xx{\times}
\def\uno{\mbox{1 \kern-.59em {\rm l}}}

\def\trp{^{\top}}
\def\inv{^{-1}}
\def\dag{{^{\dagger}}}
\def\pr{^{\prime}}

\def\rar{\rightarrow}
\def\lar{\leftarrow}
\def\lrar{\leftrightarrow}

\newcommand{\0}{\,\!}      %this is just NOTHING!
\def\one{1\!\!1\,\,}
\def\im{\imath}
\def\jm{\jmath}

\newcommand{\tr}{\mbox{tr}}
\newcommand{\slsh}[1]{/ \!\!\!\! #1}

\def\vac{|0\rangle}
\def\lvac{\langle 0|}

\def\hlf{\frac{1}{2}}
\def\ove#1{\frac{1}{#1}}

\def\Box{\square}
\def\ZZ{\mathbb{Z}}
\def\CC#1{({\bf #1})}
\def\bcomment#1{}
\def\bfhat#1{{\bf \hat{#1}}}
\def\VEV#1{\left\langle #1\right\rangle}
\def\vev#1{\langle{#1}\rangle}

\newcommand{\ex}[1]{{\rm e}^{#1}} \def\ii{{\rm i}}

\def\rr{{\rm r}} \def\rs{{\rm s}}\def\rv{{\rm v}}
\def\ri{{\rm i}}\def\rj{{\rm j}}

\newcommand{\lrbrk}[1]{\left(#1\right)}
\newcommand{\sfrac}[2]{{\textstyle\frac{#1}{#2}}}

\font\mybb=msbm10 at 12pt
\def\bb#1{\hbox{\mybb#1}}

\font\myBB=msbm10 at 18pt
\def\BB#1{\hbox{\myBB#1}}

\begin{document}

%\hfill{ }

\hfill{ hep-th/0404072}

\vspace{20pt}

\begin{center}

{\Large \bf Tree Amplitudes in Gauge Theory}

{\Large \bf  as Scalar MHV Diagrams}

\vspace{30pt}

{\bf George Georgiou and Valentin V.~Khoze}

\medskip

{\small \em
Centre for Particle Theory,
Department of Physics and IPPP,\\
University of Durham, Durham, DH1 3LE, UK
}

\vspace{10pt}

{\sffamily \tt george.georgiou,valya.khoze@durham.ac.uk }

\vspace{30pt}

{\bf Abstract}

\end{center}

It was proposed in hep-th/0403047 that all
tree amplitudes in pure Yang-Mills theory can be constructed from
known MHV amplitudes.
We apply this approach for calculating
tree amplitudes of gauge fields and fermions and find
agreement with known results.
The formalism amounts to an effective scalar perturbation theory which
offers a much simpler alternative to the usual Feynman diagrams in gauge theory
and can be used for deriving new simple expressions for tree amplitudes.
At tree level the formalism works
in a generic gauge theory, with or without supersymmetry,
and for a finite number of colours.

\vspace{0.5cm}

\setcounter{page}{0}
\thispagestyle{empty}
\newpage

%%%%%%%%%%%%%% ordinary document (end) %%%%%%%%%%%%%%%%%%%%%%%%%%%%%%%%
%\parindent=0pt
\baselineskip 6mm

\section{Introduction}

In a recent paper \cite{Witten} Witten outlined a construction which interprets
perturbative amplitudes of conformal $\cN=4$ supersymmetric gauge theory
as D-instanton contributions in a topological string theory
in twistor space.
Motivated by this correspondence, Cachazo, Svrcek and Witten \cite{CSW}
proposed a remarkable new approach for calculating
all tree-level amplitudes of $n$ gluons.
In this approach tree amplitudes in a pure gauge theory
are found by summing tree-level scalar Feynman diagrams with
new vertices.
The building blocks of this formalism are
scalar propagators $1/p^2,$ and
tree-level
maximal helicity violating (MHV) amplitudes,
which are
interpreted as new scalar vertices.
%The MHV vertices already contain an arbitrary number of gluon lines,
%and are known functions \cite{PT,BG} of kinematic variables.
Using multi-particle amplitudes as effective vertices
enables one to save dramatically on a number of permutations in usual
Feynman diagrams.

The new perturbation theory involves scalar diagrams since
MHV vertices are scalar quantities. They are linked together
by scalar propagators at tree-level, and the internal lines are continued
off-shell in a particular fashion. The final result
for any particular amplitude can be shown to be
Lorentz-covariant and is independent of a particular choice for the off-shell
continuation. The authors of \cite{CSW} derived new expressions
for a class of tree amplitudes with three negative helicities
and any number of positive ones.
It has been verified in \cite{CSW} and \cite{Zhu}
that the new scalar graph approach agrees with a number
of known conventional results for scattering
amplitudes in pure gauge theory

The motivation of this note is to apply the new diagrammatic approach of \cite{CSW}
to tree amplitudes which involve fermion fields as well as gluons.
In the presence of fermions there are two new classes of MHV vertices,
which involve one and two quark-antiquark lines. This is in addition to the
the single class of purely gluonic MHV vertices considered in \cite{CSW}.
All three classes
of vertices can in principle
be connected to one another via propagators at tree level.
This leads to new diagrams and provides us with useful
tests of the method.
Confirmation of the new diagrammatic approach of \cite{CSW} in more
general settings is important for two reasons.
First, as mentioned earlier, this approach
offers a much simpler alternative to the usual Feynman diagrams in gauge theory
and can be used for deriving a variety of new
closed-form expressions for multi-parton tree amplitudes. Of course,
in practice, in deriving multi-parton amplitudes there is no need to
calculate Feynman diagrams directly as
there are other powerful techniques based on the recursion relations
\cite{BG,Kosower}. We also note that scalar graphs as a powerful method
for calculating amplitudes in field theories with gauge fields
and fermions was introduced already in \cite{Siegel}.\footnote{In the approach of
\cite{Siegel} one also works in the helicity basis and uses scalar propagators
and scalar vertices. However the scalar vertices utilized in \cite{Siegel}
are not the MHV vertices
used in \cite{CSW} and here. We thank Warren Siegel for drawing our attention to
Ref. \cite{Siegel}.}

Our second reason for studying and generalising this approach is
its relation to string theory in twistor space.
On the string side, the SYM amplitude is interpreted
in \cite{Witten} as coming from a D-instanton
of charge $d$, where $d$ is equal to the number of negative helicity particles
minus $1$ plus the number of loops. The new scalar graph method of \cite{CSW}
is interpreted on the string side as the contribution coming entirely
from $d$ single instantons. On the other hand,
in an interesting recent paper \cite{RSV} it is argued
that the SYM amplitude is fully
determined by the opposite extreme case --
a single $d$-instanton. In principle,
there are also contributions from a mixed set of connected and disconnected
instantons of total degree $d$.

{}From the gauge theory perspective, there are two questions we can ask:

(1) does the scalar formalism of \cite{CSW} correctly incorporate
 gluinos in a generic supersymmetric theory, and

(2) does it work for diagrams with fundamental quarks in a non-supersymmetric
$SU(N)$ theory, i.e. in QCD?

It is often stated in the literature that any gauge theory is
supersymmetric at tree level.
This is because at tree level
superpartners cannot propagate in loops.
This observation, on its own, does not answer the question of how to
relate amplitudes with quarks to amplitudes with gluinos.
The colour
structure of these amplitudes is clearly
different.\footnote{Also, amplitudes with gluons and gluinos
are automatically planar at tree level. This is not the case for tree diagrams
with quarks, as they do contain $1/N$-suppressed terms in $SU(N)$ gauge theory.}
However, we will see that
the purely kinematic parts of these amplitudes are the same at tree level.

In the next section we briefly recall well-known results about decomposition of
full amplitudes into the colour factor $T_n$ and the purely kinematic partial
amplitude $A_n$.
A key point in the approach of \cite{CSW} and also in \cite{Witten,RSV} is that
only the kinematic amplitude $A_n$ is evaluated directly.
Since $A_n$ does not contain colour factors, it is the same for
tree amplitudes involving quarks and for those with gluinos.
There is an important point we should stress here.
Apriori, when comparing
kinematic amplitudes in a non-supersymmetric and in a supersymmetric theory,
we should make sure that both theories have a similar field content.
In particular, when comparing kinematic amplitudes in QCD and in SYM,
(at least initially)
we need to restrict to the SYM theory with vectors, fermions and no scalars.
Scalars are potentially dangerous, since they can propagate in the internal lines
and spoil the agreement between the amplitudes.
Hence,
while the kinematic tree-level amplitudes in massless QCD agree with those
in $\cN=1$ pure SYM, one might worry that the agreement will be lost
when comparing QCD with
 $\cN=4$ (and $\cN=2$) theories. Fortunately, this is not the case,
the agreement between amplitudes in QCD and amplitudes
in SYM theories does not depend on $\cN$.
The main point here is that in $\cN=2$ and $\cN=4$ theories, the scalars
$\phi$
couple to gluinos $\Lambda^A$ and $\Lambda^B$ from different
$\cN=1$ supermultiplets,
\be
S_{\rm Yukawa} =\ g_{\rm YM}\, \tr \, \Lambda^{-}_A [\phi^{AB}, \Lambda^{-}_B] \, + \,
g_{\rm YM}\, \tr \, \Lambda^{A+} [\overline{\phi}_{AB}, \Lambda^{B+}] \ ,
\ee
where $A,B=1,\ldots \cN $, and $\phi^{AB}=-\phi^{BA}$, hence $A\neq B$.
At the same time, in the kinematic amplitudes
quarks are identified with gluinos of the same fixed $A$, i.e.
\newline $q \leftrightarrow\Lambda^{A=1\, +}$,
$\overline{q} \leftrightarrow\Lambda^{-}_{A=1}.$

QCD-amplitudes with $m$ quarks, $m$ antiquarks and $l$
gluons in
external lines correspond to SYM-amplitudes with $m$ gluinos
$\Lambda^{1+}$, $m$ anti-gluinos $\Lambda^{-}_1$,
and $l$ gluons. Since all external (anti)-gluinos
are from the same $\cN=1$ supermultiplet,
they cannot produce scalars in the internal lines of tree diagrams. These
diagrams are the same for all $\cN=0,\ldots,4$. Of course, in
$\cN=4$ and $\cN=2$ theories there are other classes of diagrams with
gluinos from different $\cN=1$ supermultiplets, and also with scalars in external lines.
Applications of the scalar graph approach to
these more general classes of tree amplitudes in $\cN=2,4$
will be presented in \cite{WIP}.

We conclude that,
if the new formalism gives correct results for partial
amplitudes $A_n$ in a supersymmetric theory,
it will also work in a nonsupersymmetric case,
and for a finite number of colours.
Full amplitudes are then determined uniquely from the kinematic part $A_n$,
and the known expressions for $T_n$, given in \eqref{twohalf}, \eqref{cfqq}
below.
This means that for tree amplitudes questions (1) and (2) are essentially the same.

In section 3 we explain how the diagrammatic approach of \cite{CSW}
works for calculating scattering amplitudes of gluons and fermions
at tree level. This method leads to explicit and relatively simple
expressions for many amplitudes.
As a first example, using the scalar graph approach, we derive an
expression for non-MHV $- - - + \ldots +$ amplitudes
$A_n$ with two fermions and $n-2$ gluons.
We furthermore derive a non-MHV $n$-point amplitude
which involves four fermions.
These new results are checked
successfully against some previously known expressions for $n=4,5$.

In section 4 we outline
some obvious conceptual difficulties in addressing loop contributions
to $n$-point amplitudes in a massless gauge theory without an infrared cutoff.

\section{Tree Amplitudes}

We concentrate on tree-level amplitudes in a gauge theory with an arbitrary
finite number of colours.
For definiteness we take the gauge group to be $SU(N)$
and consider tree-level scattering amplitudes with arbitrary numbers of external gluons
and fermions (it is also straightforward to include scalar fields, but we leave them out
from most of what follows for simplicity). $SU(N)$ is unbroken and all fields are
taken to be massless, we refer to them generically as gluons, gluinos and quarks,
though the gauge theory is not necessarily assumed to be supersymmetric.

\subsection{Colour decomposition}

It is well-known that a full n-point amplitude ${\cal M}_n $
can be represented as a sum of products of colour factors $T_n$
and purely kinematic partial
amplitudes $A_n$,
\be
{\cal M}_n (\{k_i,h_i,c_i\}) \,=\, \sum_{\sigma} \,
T_n (\{c_{\sigma(i)}\}) \, A_n (\{k_{\sigma(i)},h_{\sigma(i)}\}) \, .
\label{one}
\ee
Here $\{c_i\}$ are colour labels of external legs $i=1 \ldots n$, and
the kinematic variables $\{k_i,h_i\}$ are on-shell external momenta and helicities:
all $k_i^2=0,$ and $h_i=\pm 1$ for gluons, $h_i=\pm {1\over 2}$ for fermions.
The sum in \eqref{one} is over appropriate simultaneous permutations $\sigma$ of
colour labels $\{c_{\sigma(i)}\}$ and kinematic variables
$\{k_{\sigma(i)},h_{\sigma(i)}\}$.
The colour factors $T_n$ are easy to determine,
and the non-trivial information about the full amplitude
${\cal M}_n $ is contained in the purely kinematic part $A_n$. If the
partial amplitudes $A_n(\{k_i,h_i\})$ are known for all permutations $\sigma$ of
the kinematic variables, the full amplitude ${\cal M}_n $ can be determined from
\eqref{one}.

We first consider tree amplitudes with arbitrary numbers of gluons
and gluinos (and with no quarks). The colour variables $\{c_i\}$ correspond to the
adjoint representation indices, $\{c_i\}=\{a_i\},$ and the colour factor $T_n$
is a single trace of generators,
\be
{\cal M}_n^{\rm tree} (\{k_i,h_i,a_i\})\, =\, \sum_{\sigma}
\tr({\rm T}^{a_{\sigma(1)}} \ldots {\rm T}^{a_{\sigma(n)}})\
A_n^{\rm tree} (k_{\sigma(1)},h_{\sigma(1)}, \ldots, k_{\sigma(n)},h_{\sigma(n)}) \, .
\label{two}
\ee
Here the sum is over $(n-1)!$ noncyclic inequivalent permutations of $n$
external particles. The single-trace structure in \eqref{two},
\be
\label{twohalf}
T_n\, = \, \tr({\rm T}^{a_{1}} \ldots {\rm T}^{a_{n}}) \, ,
\ee
implies
that all tree level amplitudes
of particles transforming in the adjoint
representation of $SU(N)$ are planar. This is not the case
neither for loop amplitudes, nor for tree amplitudes involving fundamental quarks.

Fields in the fundamental representation
couple to the trace $U(1)$ factor of the $U(N)$ gauge group.
In passing to the $SU(N)$ case this introduces power-suppressed $1/N^p$
terms. However, there is a remarkable
simplification for tree diagrams involving fundamental quarks: the factorisation property
\eqref{one} still holds. More precisely, for a fixed colour ordering $\sigma$, the amplitude
with $m$ quark-antiquark pairs and $l$ gluons (and gluinos)
is still a perfect product,
\be
T_{l+2m} (\{c_{\sigma(i)}\}) \ A_{l+2m} (\{k_{\sigma(i)},h_{\sigma(i)}\}) \, ,
\label{three}
\ee
and all $1/N^p$ corrections
to the amplitude are contained in the first term. For tree amplitudes
the exact colour factor in
\eqref{three} is \cite{MP}
\be
T_{l+2m} \, = \,
{(-1)^p\over N^p} ({\rm T}^{a_1} \ldots {\rm T}^{a_{l_1}})_{i_1 \alpha_1}
({\rm T}^{a_{l_1+1}} \ldots {\rm T}^{a_{l_2}})_{i_2 \alpha_2} \ldots
({\rm T}^{a_{l_{m-1}+1}} \ldots {\rm T}^{a_{l}})_{i_m \alpha_m} \, .
\label{cfqq}
\ee
Here $l_1, \ldots , l_{m}$ correspond to an arbitrary partition of an arbitrary
permutation of the $l$ gluon indices; $i_1, \ldots i_m$ are
colour indices of quarks, and $\alpha_1, \ldots \alpha_m$ -- of the antiquarks.
In perturbation theory each external quark is connected by a fermion line to an external
antiquark (all particles are counted as incoming). When quark $i_k$ is connected by a fermion
line to antiquark $\alpha_k$, we set $\alpha_k=\bar{i_k}$. Thus, the set of
$\alpha_1, \ldots \alpha_m$ is a permutation of the set
$\bar{i_1}, \ldots \bar{i_m}$. Finally,
the power $p$ is equal to the number of times $\alpha_k=\bar{i_k}$ minus 1.
When there is only one quark-antiquark pair, m=1 and p=0. For a general $m$,
the power $p$ in \eqref{cfqq} varies from $0$ to $m-1$.

The kinematic amplitudes $A_{l+2m}$ in \eqref{three} have the colour information stripped off
and hence do not distinguish between fundamental quarks and adjoint gluinos.
Thus,
\be
A_{l+2m}(q,\ldots,\bar{q},\ldots, g^+,\ldots, g^-,\ldots)\, = \,
A_{l+2m}(\Lambda^+,\ldots,\Lambda^-,\ldots, g^+,\ldots, g^-,\ldots)\, ,
\label{four}
\ee
where $q$, $\bar{q}$, $g^{\pm}$, $\Lambda^{\pm}$ denote
quarks, antiquarks, gluons and gluinos of $\pm$ helicity.

In section 3 we will use
the scalar graph formalism of \cite{CSW} to evaluate the kinematic
amplitudes $A_n$ in \eqref{four}.
Full amplitudes can then be determined uniquely from the kinematic part $A_n$,
and the known expressions for $T_n$ in \eqref{twohalf} and \eqref{cfqq}
by summing over the inequivalent colour orderings in \eqref{one}.

{}From now on we concentrate on the purely kinematic part of the amplitude,
$A_n$.

\subsection{Helicity amplitudes}

We will be studying tree level partial amplitudes $A_n=A_{l+2m}$
with $l$ gluons and $2m$ fermions in the helicity basis.
All external lines are defined to be
incoming, and a fermion of helicity $+{1\over 2}$ is
always connected
by a fermion propagator to a helicity $-{1\over 2}$
fermion,\footnote{This is generally correct
only in theories without scalar fields. In the $\cN=4$ theory,
a pair of positive helicity fermions, $\Lambda^{1+}$, $\Lambda^{2+}$,
can be connected to another pair of positive helicity
fermions, $\Lambda^{3+}$, $\Lambda^{4+}$, by a scalar propagator.
As already mentioned in the introduction,
for all amplitudes considered in this paper we will take external fermions
from the same $\cN=1$ supermultiplet, i.e. $A=1$, and this will rule out contributions
of scalar fields even in the $\cN=4$ theory.}
hence the number
of fermions $2m$ is always even.

A tree amplitude $A_n$ with $n$ or $n-1$ particles of positive helicity
vanishes identically. The same is true for $A_n$ with $n$ or $n-1$ particles
of negative helicity. First nonvanishing amplitudes contain $n-2$ particles
with helicities of the same sign and are called maximal helicity violating
(MHV) amplitudes.

The spinor helicity formalism\footnote{This formalism was used for
calculating scattering amplitudes first in
\cite{Berends,PT,BG}. We follow conventions of \cite{Witten} and
refer the reader also to comprehensive  reviews \cite{MP,Dixon}
where more detail and references
can be found.
Our helicity spinor conventions are summarised in the Appendix}
is defined in terms of two commuting
spinors of positive and negative chirality,
$\lambda_a$ and $\tilde\lambda_{\dot a}$.
Using these spinors, any on-shell momentum
of a massless particle, $p_\mu p^\mu=0,$ can be written as
\be
p_{a \dot a} =\ p_\mu \sigma^\mu =\ \lambda_a\tilde\lambda_{\dot a} \ .
\ee
Spinor inner products are introduced as
\be
\langle \lambda,\lambda'\rangle = \ \epsilon_{ab}\lambda^a\lambda'{}^b
 \ , \qquad
[\tilde\lambda,\tilde\lambda'] =\ \epsilon_{\dot a\dot b}
\tilde\lambda^{\dot a}\tilde\lambda'{}^{\dot b} \ .
\ee
Then a scalar product of two null vectors,
$p_{a\dot a}=\lambda_a \tilde\lambda_{\dot a}$ and
$q_{a\dot a}=\lambda'_a\tilde\lambda'_{\dot a}$, is
\be \label{scprod}
p_\mu q^\mu =\ {1\over 2}
\langle\lambda,\lambda'\rangle[\tilde\lambda,\tilde\lambda'] \ .
\ee
Momentum conservation in an $n$-point amplitude provides
another useful identity
\be
\sum_{i=1}^n \, \vev{\lambda_r ~\lambda_i}
[\tilde{\lambda}_i ~\tilde{\lambda}_s] = \ 0 \ ,
\ee
for arbitrary $1\le r,s \le n$.

In the usual perturbative evaluation of amplitudes, external on-shell lines
in Feynman diagrams are multiplied by wave-function factors:
a polarization vector $\varepsilon_{\pm}^\mu$ for each external gluon
$A_\mu$, and
spinors $u_{\pm}$ and $\bar{u_{\pm}}$ for external quarks and antiquarks.
The resulting amplitude is a Lorentz scalar.
The spinors $\lambda$ and $\tilde{\lambda}$ are precisely the wave-functions
of fermions and corresponding antifermions (see Appendix for more detail)
\be \label{wvs}
u_+ (k_i)_a =\ \lambda_{i\, a}\ , \qquad
\overline{u_+ (k_i)}_{\dot a} = \ \tilde{\lambda}_{i\, \dot a}\ ,
\ee
and the polarization vectors $\varepsilon_{\pm}^\mu$ are also defined
in a natural way in terms of $\lambda$, $\tilde{\lambda}$ (and a
`reference' spinor), as in \cite{Witten}.

$A_n (g_1^+,\ldots, g_{r-1}^+,g_r^-,g_{r+1}^+,\ldots,g_{s-1}^+,
g_s^-, g_{s+1}^+,\ldots,g_{n}^+)$
is the `mostly plus' purely gluonic MHV amplitude
with $n-2$ gluons of positive helicity
and $2$ gluons of negative helicity in positions $r$ and $s$.
To simplify notation, from now on we will not indicate the positive
helicity gluons in the mostly plus amplitudes and the negative helicity gluons in
the mostly minus amplitudes.
Also, the mostly plus maximal helicity violating amplitudes
will be referred to simply as the MHV amplitudes, and the
mostly minus maximal helicity violating amplitudes will
be called the $\overline{\rm MHV}$.
Finally, in all the amplitudes $A_n$ we will suppress the common
momentum conservation factor
of
\be
i g_{\rm YM}^{n-2} \, (2\pi)^4 \, \delta^{(4)} \big(\sum_{i=1}^n \lambda_{i a}
\tilde{\lambda}_{i \dot a} \big)
\ee
Using these conventions, the MHV gluonic amplitude is
\be
A_n (g_r^-,g_s^-)=\
{\langle\lambda_r,\lambda_s\rangle^4\over
\prod_{i=1}^n\langle\lambda_i, \lambda_{i+1}\rangle }
\equiv \
{\vev{r~s}^4 \over \prod_{i=1}^n \vev{i~i+1}} \ ,
\label{mpng}
\ee
where $\lambda_{n+1} \equiv \lambda_1$.
The corresponding
$\overline{\rm MHV}$ amplitude with positive helicity gluons
in positions $r$ and $s$ is
\be
A_n (g_r^+,g_s^+)=\
{[\tilde\lambda_r,\tilde\lambda_s]^4\over
\prod_{i=1}^n [\tilde\lambda_i, \tilde\lambda_{i+1}] }
\equiv \
{[r~s]^4 \over \prod_{i=1}^n [i~i+1]} \ .
\label{mmng}
\ee
The closed-form expressions \eqref{mpng}, \eqref{mmng}
were derived in \cite{PT,BG}.
These and other results in the helicity formalism are reviewed in
\cite{MP,Dixon}.

An MHV amplitude $A_n=A_{l+2m}$
with $l$ gluons and $2m$ fermions (from the same $\cN=1$ supermultiplet)
exists only for $m=0,1,2$.
This is because it must have precisely $n-2$ particles with positive and
$2$ with negative helicities, and our fermions always come in pairs
with helicities $\pm {1\over 2}$.
Hence, including \eqref{mpng}, there are
three types of MHV tree amplitudes,
\be
\label{threecls}
A_n (g_r^-,g_s^-) \ , \quad
A_n (g_t^-,\Lambda_r^-,\Lambda_s^+)\ , \quad
A_n (\Lambda_t^-,\Lambda_s^+,\Lambda_r^-,\Lambda_q^+) \ .
\ee
Expressions for all three MHV amplitudes in \eqref{threecls}
can be simply read off the $\cN=4$ supersymmetric formula
of Nair \cite{Nair}:
\be
A_n^{\cN=4} =\
\delta^{(8)} \big(\sum_{i=1}^n \lambda_{i a}
\eta^A_i \big)\
{1 \over \prod_{i=1}^n \vev{i~i+1}} \ ,
\label{nair}
\ee
where $\eta^A_i$, $A=1,2,3,4$ is the $\cN=4$ Grassmann coordinate.
Taylor expanding \eqref{nair} in powers of $\eta_i$, one can identify
each term in the expansion with a particular tree-level MHV amplitude
in the $\cN=4$ theory.  $(\eta_i)^k$ for $k=0,\ldots,4$ is interpreted as
the $i^{\rm th}$ particle with helicity $h_i=1-{k\over 2}$.
Hence, $h_i = \{1,{1\over 2},0,-{1\over 2},-1\},$ where zero is the
helicity of a scalar field.
It is straightforward to write down a general rule for associating a power of
$\eta$ with all component fields in $\cN=4$,
\SP{ \label{nrules}
&g^{-}_i\ \leftarrow\ \eta_i^1 \eta_i^2 \eta_i^3 \eta_i^4 \ , \quad
\phi^{AB} \ \leftarrow\  \eta_i^A \eta_i^B \ , \quad
\Lambda^{A+} \ \leftarrow\  \eta_i^A \ , \quad
g^{-}_i\ \leftarrow\  1  \ , \\
&\Lambda^{-}_{1} \ \leftarrow\  -\,\eta_i^2 \eta_i^3 \eta_i^4 \ , \quad
\Lambda^{-}_{2} \ \leftarrow\  -\,\eta_i^1 \eta_i^3 \eta_i^4 \ , \quad
\Lambda^{-}_{3} \ \leftarrow\  -\,\eta_i^1 \eta_i^2 \eta_i^4 \ ,
\quad
\Lambda^{-}_{4} \ \leftarrow\  -\,\eta_i^1 \eta_i^2 \eta_i^3 \ .
}

The amplitude \eqref{mpng} can be obtained from \eqref{nair}, \eqref{nrules}
by selecting
the $(\eta_r)^4 \ (\eta_s)^4$ term; the second amplitude in \eqref{threecls}
follows an appropriate $(\eta_t)^4 (\eta_r)^3 (\eta_s)^1$ term in \eqref{nair};
and the third amplitude
in \eqref{threecls} is an
$(\eta_r)^3 (\eta_s)^1 (\eta_p)^3 (\eta_q)^1$ term.
For our calculations in addition to \eqref{mpng}
we will need expressions for MHV amplitudes with $m=1$
and $m=2$ pairs of fermions (with the same $A$).
The MHV amplitude with two external fermions and $n-2$ gluons is
\be
\label{ndcls}
A_n (g_t^-,\Lambda_r^-,\Lambda_s^+)= \
{\vev{t~r}^3\ \vev{t~s} \over \prod_{i=1}^n \vev{i~i+1}} \ , \quad
A_n (g_t^-,\Lambda_s^+,\Lambda_r^-)= \
-\ {\vev{t~r}^3\ \vev{t~s} \over \prod_{i=1}^n \vev{i~i+1}} \ ,
\ee
where the first expression corresponds to $r<s$ and the second to $s<r$
(and $t$ is arbitrary).
The MHV amplitudes with four fermions and $n-4$ gluons on external lines are
\be
\label{ndcls2}
A_n (\Lambda_t^-,\Lambda_s^+,\Lambda_r^-,\Lambda_q^+)
= \
{\vev{t~r}^3\ \vev{s~q} \over \prod_{i=1}^n \vev{i~i+1}} \ , \quad
A_n (\Lambda_t^-,\Lambda_r^-,\Lambda_s^+,\Lambda_q^+)
= \
-\ {\vev{t~r}^3\ \vev{s~q} \over \prod_{i=1}^n \vev{i~i+1}}
 \
\ee
The first expression in \eqref{ndcls2}
corresponds to $t<s<r<q,$ the second -- to $t<r<s<q,$
and there are other similar aexpressions,
obtained by further permutations of fermions, with the overall
sign determined by the ordering.

The $\overline{\rm MHV}$ amplitude can be obtained, as always,
by exchanging helicities $+\leftrightarrow -$ and
$\vev{i~j} \leftrightarrow [i~j].$
Expressions \eqref{ndcls}, \eqref{ndcls2} can also be derived from $\cN=1$ supersymmetric
Ward identities, as in \cite{MP,Dixon}.

All amplitudes following from Nair's general expression \eqref{nair}
are analytic in the sense that they depend only on $\vev{\lambda_i~\lambda_j}$
spinor products,
and not on $[\tilde\lambda_i~ \tilde\lambda_i].$ These include amplitudes
in \eqref{mpng}, \eqref{ndcls}, \eqref{ndcls2}, as well as more complicated classes
of amplitudes,
i.e. with external gluinos $\Lambda^A,$ $\Lambda^{B\neq A},$ etc, and
with external scalar fields. These extra classes of analytic amplitudes,
interpreted as vertices in the scalar graph approach, are being investigated
in \cite{WIP} and will not be discussed here further.

\section{Calculating Amplitudes Using Scalar Graphs}

The formalism of \cite{CSW}
represents all non-MHV tree amplitudes (including $\overline{\rm MHV}$)
as sums of tree diagrams in an effective scalar perturbation theory.
The vertices in this theory are the MHV amplitudes \eqref{threecls},
continued off-shell as described below, and connected by scalar
bosonic propagators $1/p^2$.

An obvious question one might ask is why one should use
the $1/p^2$ propagator when connecting fermion lines in MHV vertices \eqref{threecls}.
To answer this, recall that the vertices \eqref{threecls} already contain
the wave-function factors for all external lines, including fermions \eqref{wvs}.
An incoming fermion in one MHV vertex, connected by $1/p^2$ to an
incoming antifermion of another MHV vertex,
corresponds to a factor of
\be \label{fermprop}
u_+ (p)_a \ {1\over p^2} \ \overline{u_+ (p)}_{\dot a} =\
{{{p\!\!\!/}\;}_{a \dot a} \over p^2} \ ,
\ee
for an internal line in the usual Feynman diagram, which is just the
right answer. There is a subtlety with choosing the ordering
of fermions in each vertex which is explained in Appendix in
equations \eqref{comprelcor} and \eqref{Apthreeef}.

When one leg of an MHV vertex is connected by a propagator
to a leg of another MHV vertex, both of these legs become internal
to the diagram and have to be continued off-shell. Off-shell continuation
is defined as follows \cite{CSW}:
we pick an arbitrary spinor $\eta^{\dot a}$ and define
$\lambda_a$ for any internal line carrying momentum $p_{a\dot a}$
by
\be \label{ofsh}
\lambda_a=p_{a\dot a}\eta^{\dot a}\ .
\ee
The
same $\eta$ is used for all the off-shell lines in all diagrams
contributing to a given amplitude.
In practice it will be convenient to choose
$\eta^{\dot a}$ to be equal to $\tilde{\lambda}^{\dot a}$
of one of the external legs of negative helicity.
External lines in a diagram remain on-shell, and for them
$\lambda$ is defined in the usual way.

Since in each MHV vertex \eqref{threecls} there are precisely two lines
with negative helicities, and since a propagator always connects lines
with opposite helicities, there is a simple relation between the number
of negative helicity particles in a given amplitude and the number of
MHV vertices needed to construct it,
\be
q_{(-1)}+ q_{(-{1\over 2})} = \ \sum v -1 \ .
\ee
Here $q_{(-1)}$ is the number of negative helicity gluons,
$q_{(-{1\over 2})}$ is the number of negative helicity fermions,
and $\sum v$ is the total number of all MHV vertices \eqref{threecls}
needed to construct this amplitude.

This formalism leads to explicit and relatively simple
expressions for many amplitudes. $n$-point amplitudes
with three particles of negative helicity is the next case beyond
simple MHV amplitudes.

\subsection{Calculating $---+++\ldots ++$ amplitudes with 2 fermions}

To illustrate the power of the method in pure gauge theory,
the authors of \cite{CSW} have calculated $n$ gluon amplitudes
with three consecutive gluons of negative helicity $---+++\ldots++$.
In order to see precisely what is new when fermions are present,
and to provide another useful application of the method,
in this section we will
calculate a similar amplitude with three negative helicities
which are now carried by a fermion and two gluons.

We consider an $n$-point amplitude,
\be \label{case1}
A_n (\Lambda_1^-,g_2^-,g_3^-,\Lambda_k^+) \ ,
\ee
with one fermion
and two gluons of negative helicities consecutive to each other, and
a positive-helicity fermion in arbitrary position $k$, such that
$3<k\le n.$ As in the case considered in \cite{CSW}, this amplitude comes from
scalar diagrams with two vertices and one propagator, but in our case
there is more that one type of
vertex in \eqref{threecls}. There are three classes
of scalar diagrams which contribute to $A_n (\Lambda_1^-,g_2^-,g_3^-,\Lambda_k^+),$
they are depicted in Figure 1. First two classes of diagrams involve the
first and the second vertex in \eqref{threecls}, and the third class involves
two second vertices in \eqref{threecls}.
\begin{figure}[ht]
\label{fig1}
\psfrag{i+}{\large$i\,+$}
\psfrag{i+1}{\large$(i+1)\,+$}
\psfrag{k+}{\large$k\,+$}
\psfrag{n+}{\large$n\,+$}
\psfrag{n}{\large$n\,+$}
\psfrag{1-}{\large$1\,-$}
\psfrag{2-}{\large$2\,-$}
\psfrag{2}{\large$2\,-$}
\psfrag{3}{\large$3\,-$}
\psfrag{4}{\large$4\,+$}
\psfrag{+}{\large$+$}
\psfrag{-}{\large$-$}
\begin{center}
{\scalebox{0.8}{
\includegraphics{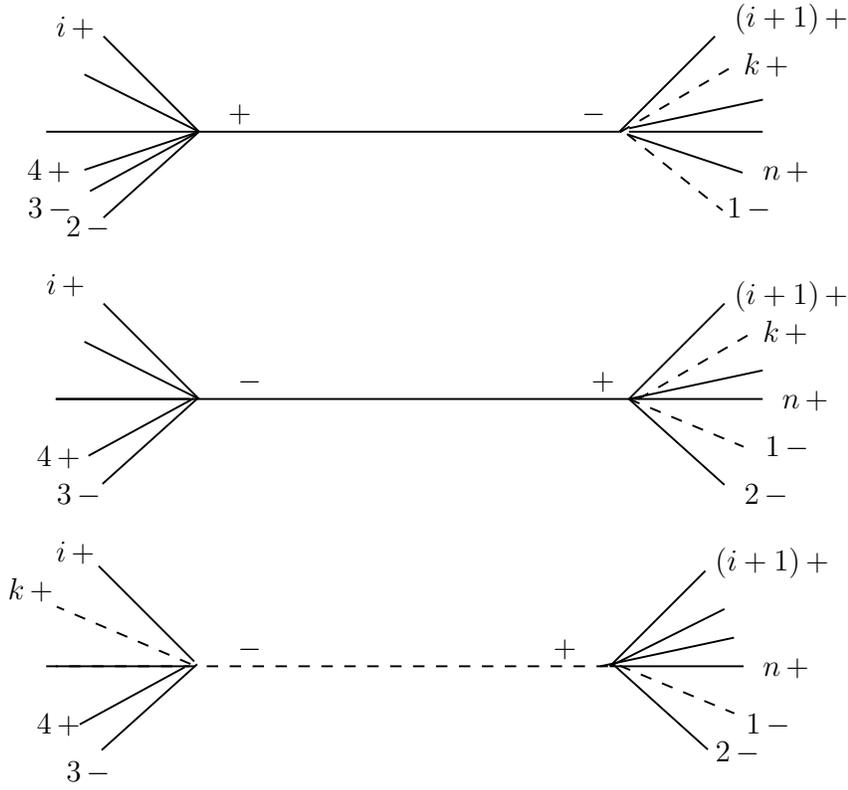}}
}
\end{center}
\caption{\it Tree diagrams with MHV vertices contributing to the
$---+++\ldots ++$ amplitude with 2 fermions and $n-2$ gluons in
Eq.~\eqref{case1}.
Fermions are represented by dashed lines and
gluons -- by solid lines.}
\end{figure}

First two diagrams in Figure 1 involve a gluon exchange.
The third diagram
involves a fermion exchange, and
can be schematically represented as
\be \label{Apthrnew}
A(\Lambda_1^-,g_2^-,\underline{\Lambda_{I}^+})
 \  {1 \over q^2_{I}}\
 A(g_3^-,\Lambda_k^+,\underline{\Lambda_{-I}^-})
\ .
\ee
Here $\underline{\Lambda_{I}^+}$ and $\underline{\Lambda_{-I}^-}$
denote internal off-shell fermions, which are Wick-contracted
via the scalar propagator $1/q^2_{I}.$
The order in which these internal fermion appear in \eqref{Apthrnew}
is according to the ket$^+$ ket$^-$ prescription discussed in the Appendix.

The three diagrams in Figure 1 give
\SP{ \label{mmmPPPP}
A_n &=\sum_{i=3}^{k-1} {\vev{1~ (2,i)}^2 \ \vev{k~ (2,i)}
\over
\vev{(2,i)~i+1}\vev{i+1~i+2} \ldots \vev{n~1}}\
{1\over q_{2i}^2}\
{\vev{2~3}^3 \over \vev{(2,i)~2} \vev{3~4}\ldots
\vev{i~(2,i)}} \\
&+\sum_{i=4}^{k-1} {\vev{1~2}^2 \ \vev{k~2}
\over \vev{2~(3,i)}
\vev{(3,i)~i+1} \ldots \vev{n~1}}\  {1\over q_{3i}^2}\
{\vev{(3,i)~3}^3 \over  \vev{3~4} \ldots \langle
i-1~i\rangle\vev{i~(3,i)}} \\
&+\sum_{i=k}^n {-\vev{1~2}^2 \over
\vev{(3,i)~i+1}\vev{i+1~i+2} \ldots \vev{n~1}}\ {1\over q_{3i}^2}\
{\vev{(3,i)~3}^2\ \vev{k~3}
\over  \vev{3~4} \ldots \langle
i-1~i\rangle\vev{i~(3,i)}} \ .}
Following notations of \cite{CSW} we have introduced
$q_{ij}=p_i+p_{i+1}+\dots+p_j.$ The corresponding
off-shell spinor
$\lambda_{ij\,a}$ is defined as in \eqref{ofsh},
$\lambda_{ij\,a}=q_{ij\,a\dot a}\eta^{\dot a}$.
All other spinors $\lambda_i$ are on-shell and
$\vev{i~(j,k)}$ is an abbreviation for a spinor product
$\vev{\lambda_i,\lambda_{jk}}$.

As in \cite{CSW} we choose $\eta^{\dot a}=\tilde\lambda_2^{\dot a}$,
and evaluate the amplitude \eqref{mmmPPPP} as a function
of the on-shell kinematic variables, $\lambda_i$ and
$\tilde\lambda_{1}$, $\tilde\lambda_{2}$, $\tilde\lambda_{3}$.
The final expression for the amplitude can be written as
the sum of three terms:
\be
A_n \ =\ {1 \over \prod_{l=3}^n\ \vev{l~l+1}}
\ \left[ A_n^{(1)} + A_n^{(2)} + A_n^{(3)}
\right] \ .
\label{threeA}
\ee
We have to treat the $i=3$ term in the first sum in \eqref{mmmPPPP}
and the $i=n$ term in the last sum in \eqref{mmmPPPP} separately,
as individually they are singular for our choice of
$\eta^{\dot a}=\tilde\lambda_2^{\dot a}$. These two terms are
assembled into $A_n^{(3)}$.

For $i\neq 3$ the first and the second lines in \eqref{mmmPPPP}
give
\SP{\label{Aone}
A_n^{(1)} = \ &\sum_{i=4}^{k-1}\ {\vev{i~i+1}\over
\vev{i^-|{q\!\!\!/}_{2,i}|2^-}
\vev{(i+1)^-|{q\!\!\!/}_{i+1,2}|2^-}
\vev{2^-|{q\!\!\!/}_{2,i}|2^-}}
\\
&\cdot\left(
{\vev{3~2}^3 \vev{1^-|{q\!\!\!/}_{2,i}|2^-}^2
\vev{k^-|{q\!\!\!/}_{2,i}|2^-}
\over q_{2,i}^2}
+{\vev{1~2}^2 \vev{k~2} \vev{3^-|{q\!\!\!/}_{i+1,2}|2^-}^3
\over q_{i+1,2}^2}
\right) \ .
   }
In evaluating \eqref{Aone} we used the Lorentz-invariant
combination
$\vev{ i^-|{p\!\!\!/}\;\;|j^-} =i^ap_{a\dot a}j^{\dot a},$
see Eq.~\eqref{lorinv} in the Appendix.
We also used
momentum conservation to set $q_{3,i}=-q_{i+1,2}$, and the anticommuting
nature of spinor products to simplify the formula.

The second term in \eqref{threeA}
is the contribution of the third line in \eqref{mmmPPPP}
for $i\neq n$. We find
\be
\label{Atwo}
A_n^{(2)} = \ \sum_{i=k}^{n-1}\ \vev{i~i+1}\ {1\over q_{i+1,2}^2}\
{\vev{1~2}^2\  \vev{k~3} \ \vev{3^-|{q\!\!\!/}_{i+1,2}|2^-}^2
\over
\vev{i^-|{q\!\!\!/}_{i+1,2}|2^-} \vev{(i+1)^-|{q\!\!\!/}_{i+1,2}|2^-}} \ .
\ee
The remaining terms -- the $i=3$ term in the first sum in \eqref{mmmPPPP}
and the $i=n$ term in the last sum in \eqref{mmmPPPP} -- both contain
a factor $[2~\eta]$ in the denominator and are singular for
$|\eta^-\rangle= |2^-\rangle$. For the method to work, the singularity
has to cancel between the two terms. This is indeed the case,
and rather nontrivially the cancellation occurs between the
diagrams of different types -- the first and the last in Figure 1 --
with different MHV vertices. After the singularity cancels, the
remaining finite contribution from these two terms is derived by
setting $|\eta^-\rangle= |2^-\rangle + |\epsilon^-\rangle $,
bringing two terms to a common denominator and using
Schouten's identity,
$[\alpha~\beta][\gamma~\delta]+[\alpha~\gamma][\delta~\beta]
+[\alpha~\delta][\beta~\gamma]=0$. In the end $|\epsilon^-\rangle $
is set to zero.
The result is
\be
\label{Athree}
A_n^{(3)} = \
 \vev{3~1}\vev{3~k} \left(
 {s_{13}+ 2 (s_{12} + s_{23})\over [1~2][2~3]}
 + {\vev{3~1}\vev{2~n}\over [1~2]\vev{1~n}}
+ {\vev{3~1}\vev{2~4}\over [2~3]\vev{3~4}}
- {\vev{3~1}\vev{2~k}\over [2~3]\vev{3~k}}
\right)
\ee
where $s_{km}=(p_k+p_m)^2 = \vev{k~m}[k~m]$.

\subsection{Tests of the amplitude \eqref{threeA}--\eqref{Athree}}

We will now test our result
for an $n$-point $---+++\ldots++$ amplitude with 2 fermions, against
some known simple cases with $n=4,5$.

We first consider a $4$-point amplitude with three negative helicities,
$A_4 (\Lambda_1^-,g_2^-,g_3^-,\Lambda_4^+)$ and check if this
amplitude vanishes.
Hence, we set $n=4=k$ and find that
\be
A_n^{(1)}= \ 0 \ , \qquad A_n^{(2)}=\ 0 \ ,
\ee
since both expressions, \eqref{Aone} and \eqref{Atwo},
are proportional to $\sum_{i=4}^3 \equiv 0$.
The remaining contribution $A_n^{(3)}$ in \eqref{Athree}
gives
\be
A_n^{(3)} = \
 \vev{3~1}\vev{3~k} \biggl(
 -{\vev{1~3}[1~3]\over [1~2][2~3]}
 + {\vev{3~1}\vev{2~4}\over [1~2]\vev{1~4}} \biggr)=\ 0 \ .
\ee
Here we first used that for $n=4$ case $s_{12}+s_{23}+s_{13}=0$,
and a momentum conservation identity,
$\vev{4~1}[1~3]+\vev{4~2}[2~3]=0.$

The next test involves a $5$-point amplitude with three negative helicities,
\newline
$A_5 (\Lambda_1^-,g_2^-,g_3^-,\Lambda_4^+,g_5^+)$. This is necessarily an
$\overline{\rm MHV}$ amplitude, or a mostly minus MHV amplitude
which is  (cf second equation in \eqref{ndcls})
\be
A_5 (\Lambda_1^-,g_2^-,g_3^-,\Lambda_4^+,g_5^+) = \
-\ {[5~4]^3\ [5~1]\over \prod_{i=1}^5 [i~i+1]} =\
{[4~5]^2 \over [1~2] [2~3] [3~4]} \ .
\ee
We set $n=5$, $k=4$ and evaluate expressions in \eqref{Aone}-\eqref{Athree}.
First, we notice again that
\be
A_n^{(1)}= \ 0 \ ,
\ee
since $\sum_{i=4}^3 \equiv 0$. However, $A_n^{(2)}$ and $A_n^{(3)}$ are
both non-zero,
\be
{A_n^{(2)}\over \prod_{l=3}^5\ \vev{l~l+1}}
= \ {[4~2]^2\ \vev{1~2}^2 \over \vev{1~5}^2} \
{1 \over [1~2][2~3][3~4]} \ ,
\label{antwo}
\ee
\be
{A_n^{(3)} \over \prod_{l=3}^5\ \vev{l~l+1}}
= \ {\vev{3~1} \over \vev{4~5}\vev{5~1}} \
\left(2{\vev{4~5} [4~5] \over [1~2][2~3]}
-{\vev{1~3}[1~3]\over [1~2][2~3]}
+{\vev{3~1}\vev{2~5} \over [1~2]\vev{1~5}}
\right)
\label{anthree}
\ee
We further use a momentum conservation identity to re-write
the first factor in \eqref{antwo} as
\be
\label{415}
{[4~2]^2\ \vev{1~2}^2 \over \vev{1~5}^2} =\
\left([4~5]-{\vev{1~3}[3~4] \over \vev{1~5}}\right)^2=\
[4~5]^2 +{\vev{1~3}^2\ [3~4]^2 \over \vev{1~5}^2} -
2{\vev{1~3}[3~4][4~5] \over \vev{1~5}}
\ee
Now, the last term on the right hand side of \eqref{415}
cancels the first term in brackets
in \eqref{antwo} and the second term in \eqref{415} cancels
the last two terms in \eqref{antwo} via an identity
\be
[3~4]\vev{4~5}+[3~1]\vev{1~5}+[3~2]\vev{2~5}=0 \ .
\ee
The remaining $[4~5]^2$ term on the right hand side of \eqref{antwo}
leads to the final
answer for the amplitude,
\be
{[4~5]^2 \over [1~2] [2~3] [3~4]} \ .
\ee

We expect that other, more involved tests of the amplitude
at the $6$-point level and beyond will also be successful.

\subsection{Calculating $---+++\ldots ++$ amplitudes with 4 fermions}

Since the scalar graph method gives correct results
for non-MHV amplitudes with 2 external fermions, the next step
is to apply this method to 4-fermion amplitudes.
In this section
we will calculate an $n$-point amplitude with 4 fermions for three
negative helicities consecuitive to each other,
\be \label{case2}
A_n(g_1^-,\Lambda_2^-,\Lambda_3^-,\Lambda_p^+,\Lambda_q^+) \ ,
\ee
where $3<p<q\le n.$
As always, positive-helicity gluons in amplitudes will not be indicated
explicitly, unless they appear in internal lines.

There are four scalar diagrams which contribute to this process. They
are drawn in Figure 2.
\begin{figure}[tbp]
\label{figure2}
\psfrag{i+}{\LARGE${i\,+}$}
\psfrag{1-}{\LARGE${1\,-}$}
\psfrag{2-}{\LARGE${2\,-}$}
\psfrag{3-}{\LARGE${3\,-}$}
\psfrag{p+}{\LARGE${p\,+}$}
\psfrag{q+}{\LARGE${q\,+}$}
\psfrag{i+1+}{\LARGE${(i+1)\,+}$}
\psfrag{n}{\LARGE${n\,+}$}
\psfrag{+}{\LARGE${+}$}
\psfrag{-}{\LARGE${-}$}
\begin{center}
\scalebox{0.50}{\includegraphics{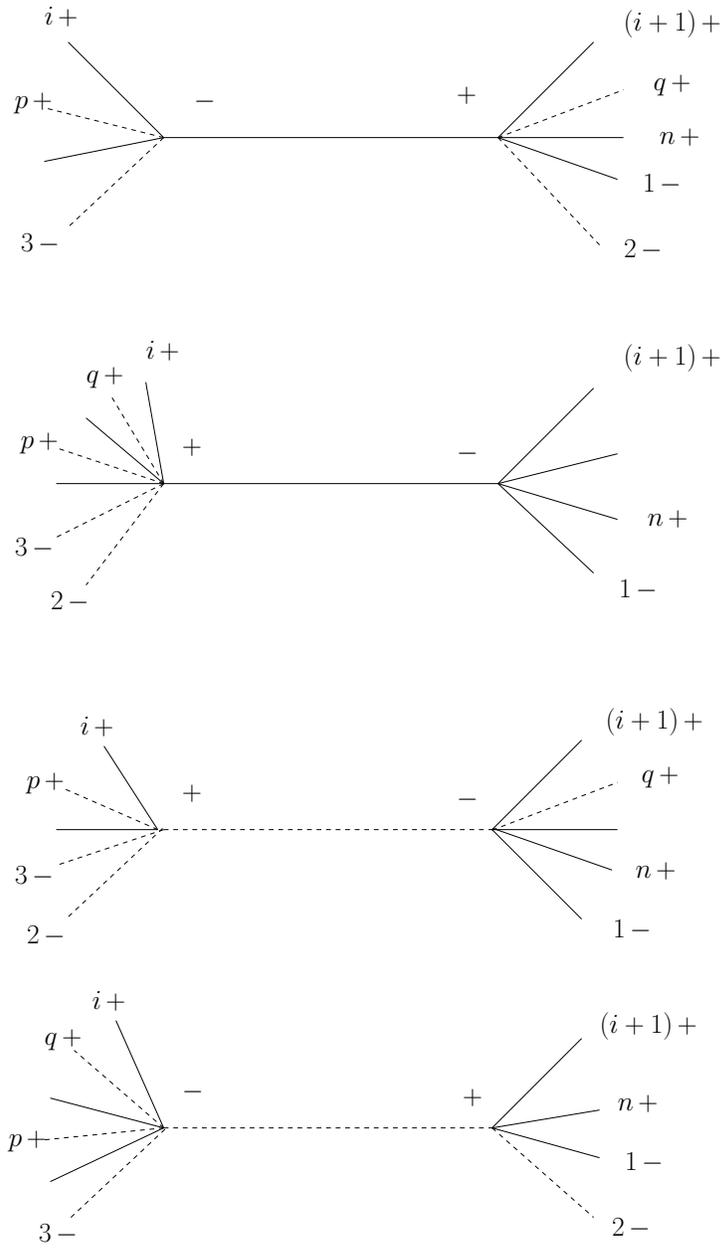}}
\end{center}
\caption{\it Diagrams contributing to the 4-fermion $n$-point amplitude \eqref{case2}
}
\end{figure}
The first diagram in Figure 2 is a gluon exchange between two
2-fermion MHV-vertices. This diagram has a schematic form,
\be \label{beone}
A(g_1^-,\Lambda_2^-,\underline{g_I^+},\Lambda_q^+) \ {1\over q_I^2} \
A(\Lambda_3^-,\Lambda_p^+,\underline{g_{-I}^-}) \ .
\ee
Here $\underline{g_I^+}$ and $\underline{g_{-I}^-}$ are off-shell (internal)
gluons which are Wick-contracted via a scalar
propagator, and $I=(3,i)$.

The second diagram involves a gluon exchange between a 0-fermion and a
4-fermion MHV vertex,
\be \label{betwo}
A(g_1^-,\underline{g_I^-}) \ {1\over q_I^2} \
A(\Lambda_2^-,\Lambda_3^-,\Lambda_p^+,\Lambda_q^+,\underline{g_{-I}^+}) \ ,
\ee
with external index $I=(2,i).$
Note that this diagram exists only for $n>5$, i.e. there must be
at least one $g^+$ in the first vertex, otherwise it is a 2-point vertex which
does not exist.

The third and the fourth diagrams in Figure 2 involve a fermion exchange
between a 2-fermion and a 4-fermion MHV vewrtices. They are given, respectively
by
\be \label{feone}
A(\Lambda_2^-,\Lambda_3^-,\Lambda_p^+,\underline{\Lambda_{-I}^+})
\ {1\over q_I^2} \
A(\Lambda_q^+,g_1^-,\underline{\Lambda_I^-}) \ ,
\ee
with $I=(2,i),$ and
\be \label{fetwo}
A(g_1^-,\Lambda_2^-,\underline{\Lambda_I^+})\ {1\over q_I^2} \
A(\Lambda_3^-,\Lambda_p^+,\Lambda_q^+,\underline{\Lambda_{-I}^-}) \ ,
\ee
with $I=(3,i).$
Both expressions, \eqref{feone} and \eqref{fetwo}, are written in the form
which is in agreement with our ordering prescription for internal
fermions, ket$^+$ ket$^-$.

We will continue using the same of-shell prescription
$\eta^{\dot a}=\tilde\lambda_2^{\dot a}$
as in the section {\bf 3.1}. But there is an importnat
simplification in the present case compared to {\bf 3.1}
-- there will be no
singular terms appearing in individual diagrams.
This is because the reference spinor $\eta^{\dot a}=\tilde\lambda_2^{\dot a}$
now corresponds to a gluino, rather than a gluon $g^-$. The reason
for singularities encountered in section  {\bf 3.1} and in Ref.~\cite{CSW} was
simply the singular collinear limit of the 3-gluon vertex where all gluons went on-shell.
We will see that these singularities would not occur
in the present case, and not having to cancel them will save us some
work.\footnote{In view of this, the calculation in section  {\bf 3.1}
could have been made simpler, if we had chosen
the reference spinor to be the spinor of the negative-helicity gluino
$\Lambda_1^-.$}

For extra clarity
we will first present a simpler version of the evaluation of \eqref{case2}
for the case $n=5.$ This result will then be generalized to all values of $n$.
For $n=5,$ we set $p=4$ and $q=5$ in \eqref{case2}, and
the calculation is straightforward.

1. The first diagram in Figure 2, Eq.~\eqref{beone}, is
\SP{ \label{one1}
{-\vev{1~2}^2 \over (\vev{2~3}[2~3]+\vev{2~4}[2~4])\vev{5~1}[2~1]}  \, &\cdot \,
{1 \over \vev{3~4}[3~4]}  \, \cdot \,
\vev{4~3}[2~4]^2 \\
&=\
{[2~4]^2\vev{1~2}^2 \over [3~4] (\vev{2~3}[2~3]+\vev{2~4}[2~4])\vev{5~1}[2~1]} \ .
}

2. The second diagram is zero.

3. The third diagram, Eq.~\eqref{feone}, is
\SP{ \label{three3}
{-\vev{2~3}^2 \over (\vev{2~3}[2~3]+\vev{2~4}[2~4])\vev{3~4}} \, &\cdot \,
{1 \over \vev{5~1}[5~1]} \, \cdot \,
{\vev{5~1}[2~5]^2 \over [2~1]} \\
&=\ {-[2~5]^2\vev{2~3}^2 \over [2~1] [5~1] (\vev{2~3}[2~3]+\vev{2~4}[2~4])\vev{3~4}} \ .
}

4. The fourth diagram, Eq.~\eqref{fetwo}:
\be
 \label{four4}
 {\vev{2~1}\over [2~1]}  \, \cdot \,
{1 \over \vev{1~2}[1~2]}
 \, \cdot \,
{\vev{3~1}^2[2~1] \over \vev{3~4}\vev{5~1}} \ = \
{\vev{3~1}^2 \over [2~1] \vev{3~4}\vev{5~1}} \ .
\ee

Now, we need to add up the three contributions.
We first combine the expressions in \eqref{one1} and
\eqref{three3} into
\be
{[4~5]^2 \over [2~1][3~4][5~1]} -
{\vev{3~1}^2 \over [2~1] \vev{3~4}\vev{5~1}}
\ee
using momentum conservation identitites, and the fact that
$\vev{2~3}[2~3]+\vev{2~4}[2~4]=-\vev{3~4}[3~4]+\vev{5~1}[5~1].$
Then, adding the remaining contribution \eqref{four4} we
obtain the final result for the amplitude,
\be
A_5(g_1^-,\Lambda_2^-,\Lambda_3^-,\Lambda_4^+,\Lambda_5^+)
\, = \,
{-[4~5]^3[2~3] \over [1~2][2~3][3~4][4~5][5~1]} \ .
\ee
which is the precisely right answer for the MHV-bar
5-point `mostly minus' diagram!

\bigskip

We now present the general expression for the amplitude with $n$
external legs.
Using the same prescription for the vertices as above the
first diagram of Figure 2 gives
\SP{
A_n^{(1)} &=\sum_{i=p}^{q-1} {\vev{1~2}^2 \ \vev{1~q}
\over
\vev{2~(3,i)}\vev{(3,i)~i+1}\vev{i+1~i+2} \ldots \vev{n~1}}\
{1\over q_{3i}^2}\
{{\vev{3~-(3,i)}}^3 \vev{p~-(3,i)}
\over
\vev{3~4} \ldots \vev{i~-(3,i)} \vev{-(3,i)~3}\ }\\
&=-{1 \over \prod_{l=3}^n\ \vev{l~l+1}}\sum_{i=p}^{q-1}
\frac{ {\vev{1~2}}^2 \vev{1~q}\vev{i~i+1}  \vev{3^-|{q\!\!\!/}_{i+1,2}|2^-}^2
\vev{p^-|{q\!\!\!/}_{2,i}|2^-}}
{q_{i+1,2}^2\vev{i^-|{q\!\!\!/}_{2,i}|2^-}
\vev{(i+1)^-|{q\!\!\!/}_{i+1,2}|2^-}
\vev{2^-|{q\!\!\!/}_{2,i}|2^-}} \ .
 }
For the second diagram of Figure 2 one obtains
\SP{
A_n^{(2)} &=\sum_{i=q}^{n-1} {\vev{1~(2,i)}^3 \
\over
\vev{(2,i)~i+1}\vev{i+1~i+2} \ldots \vev{n~1}}\
{1\over q_{2,i}^2}\
{-{\vev{2~3}}^2 \vev{p~q}
\over
\vev{3~4} \ldots \vev{i~-(2,i)} \vev{-(2,i)~2}\ }\\
&={1 \over \prod_{l=3}^n\ \vev{l~l+1}}\sum_{i=q}^{n-1}
\frac{ {\vev{2~3}}^2 \vev{p~q}\vev{i~i+1}
\vev{1^-|{q\!\!\!/}_{2,i}|2^-}^3}
{q_{2,i}^2\vev{i^-|{q\!\!\!/}_{2,i}|2^-}
\vev{(i+1)^-|{q\!\!\!/}_{i+1,2}|2^-}
\vev{2^-|{q\!\!\!/}_{2,i}|2^-}} \ .
 }
The contribution of the third diagram of Figure 2 is
\SP{
A_n^{(3)} &=\sum_{i=p}^{q-1} -{\vev{1~(2,i)}^2 \ \vev{1~q}
\over
\vev{(2,i)~i+1}\vev{i+1~i+2} \ldots \vev{n~1}}\
{1\over q_{2,i}^2}\
{-{\vev{2~3}}^2 \vev{p~-(2,i)}
\over
\vev{3~4} \ldots \vev{i~-(2,i)} \vev{-(2,i)~2}\ }\\
&={1 \over \prod_{l=3}^n\ \vev{l~l+1}}\sum_{i=p}^{q-1}
\frac{ {\vev{2~3}}^2 \vev{1~q}\vev{i~i+1} \vev{p^-|{q\!\!\!/}_{2,i}|2^-}
\vev{1^-|{q\!\!\!/}_{2,i}|2^-}^2}
{q_{2,i}^2\vev{i^-|{q\!\!\!/}_{2,i}|2^-}
\vev{(i+1)^-|{q\!\!\!/}_{i+1,2}|2^-}
\vev{2^-|{q\!\!\!/}_{2,i}|2^-}} \ .
 }
Finally, from the fourth diagram of Figure 2 we get
\SP{
A_n^{(4)} &=\sum_{i=q}^{n} {\vev{1~2}^2 \ \vev{1~(3,i)}
\over
\vev{2~(3,i)}\vev{(3,i)~i+1}\vev{i+1~i+2} \ldots \vev{n~1}}\
{1\over q_{3,i}^2}\
{-{\vev{3~-(3,i)}}^3 \vev{p~q}
\over
\vev{3~4} \ldots \vev{i~-(3,i)} \vev{-(3,i)~3}\ }\\
&=-{1 \over \prod_{l=3}^n\ \vev{l~l+1}}\sum_{i=q}^{n}
\frac{ {\vev{1~2}}^2 \vev{p~q}\vev{i~i+1}  \vev{3^-|{q\!\!\!/}_{i+1,2}|2^-}^2
\vev{1^-|{q\!\!\!/}_{2,i}|2^-}}
{q_{i+1,2}^2\vev{i^-|{q\!\!\!/}_{2,i}|2^-}
\vev{(i+1)^-|{q\!\!\!/}_{i+1,2}|2^-}
\vev{2^-|{q\!\!\!/}_{2,i}|2^-}} \ .
 }
One can combine the result for the first and the third diagram to get:
\bea
A_n^{(1,3)}={1 \over \prod_{l=3}^n\vev{l~l+1}} \sum_{i=p}^{q-1}
\frac{\vev{1~q}\vev{i~i+1}\vev{p^-|{q\!\!\!/}_{2,i}|2^-}} {\vev{i^-|{q\!\!\!/}_{2,i}|2^-}
\vev{(i+1)^-|{q\!\!\!/}_{i+1,2}|2^-}
\vev{2^-|{q\!\!\!/}_{2,i}|2^-}}\\
\cdot\left( -\frac{\vev{1~2}^2 \vev{3^-|{q\!\!\!/}_{i+1,2}|2^-}^2}{q_{i+1,2}^2}+
 \frac{\vev{2~3}^2 \vev{1^-|{q\!\!\!/}_{2,i}|2^-}^2}{q_{2,i}^2} \right) \ .
\eea
The final result for the $n$-point amplitude is with 4 fermions
\bea
\label{finr}
A_n(g_1^-,\Lambda_2^-,\Lambda_3^-,\Lambda_p^+,\Lambda_q^+) \ =\
A_n^{(1,3)}+A_n^{(2)}+A_n^{(4)} \ .
\eea
All the individual terms are regular, and the equation above
is the final result of this section.

\section{Loop Amplitudes and IR Divergencies}

An intriguing question is how to go beyond the tree level.
There are some obvious conceptual problems in trying to
work with loop amplitudes, directly in 4 dimensions
and without an infrared cutoff
(e.g. in the superconformal $\cN=4$ SYM).
Loop amplitudes in massless gauge theories suffer from severe infrared (IR)
-- soft and collinear -- divergencies.
At tree-level there are no integrations over loop momenta
and IR divergencies in the amplitudes can be avoided by selecting a
non-exceptional set of external momenta (i.e the set with none of
the external momenta being collinear or soft).
Hence tree amplitudes can be made IR finite
and it is meaningful to be calculating them
directly in 4D without an explicit IR cutoff.

Loop amplitudes,
however, are always IR divergent, in other words, one cannot
choose a set of external momenta which would make an on-shell
loop amplitude finite in 4D.
These IR divergencies simply reflect the fact that
the naive S-matrix and scattering amplitudes simply do not exist
in a gauge theory with massless particles. In QCD (and QED) this problem is
avoided either by calculating cross-sections directly or by
defining new asymptotic initial and final states with indefinite numbers
of massless quarks and gluons.
The Kinoshita-Lee-Naunberg theorem states that all the IR divergencies
cancel (collinear with collinear, soft with soft) in properly
defined physical observables,
when one sums over degenerate initial as well as final states.

In the superconformal $\N=4$ theory the situation is less clear.
Well-defined observables of this theory are not the amplitudes, but
Green functions of gauge-invariant composite operators, and
the latter were used successfully in the context of the AdS/CFT correspondence.

It would be very interesting to understand better what are
the relevant
infrared-safe quantities in a 4-dimensional
gauge theory
which can be deduced from a
dual string theory in twistor space.

\bigskip
\bigskip

\centerline{\bf Acknowledgements}

VVK would like to thank participants
of `Strings, Gauge Fields and Duality' conference in Swansea,
where this work was started,
for useful conversations. He is also grateful to
 Adrian Signer for comments.
VVK is supported in part by a PPARC Senior Research Fellowship.
GG acknowledges a grant form the State Scholarship Foundation
of Greece (I.K.Y.)

%\bigskip
\newpage

\section*{Appendix: Note on Spinor Conventions}

Spinor products are defined as
\be \label{Aspicon}
\vev{i~j} \equiv \ \vev{i^-|j^+} = \ \lambda_i^{~a} \lambda_{j~a}
 \ , \qquad
[i~j] \equiv \ \vev{i^+|j^-}  = \ \tilde\lambda_i^{~\dot a}
\tilde\lambda_{j ~\dot a} \ .
\ee
Here spinor indices are raised and lowered with $\epsilon$-symbols,
and we follow the sign conventions of \cite{Witten,CSW}.
It should be noted, that
slightly different sign conventions from \eqref{Aspicon}
have been used in earlier literature for $[i~j]$. For example,
in \cite{Dixon} the dotted spinor product, $[i~j],$ is defined
as $\tilde\lambda_{i~\dot a} \, \tilde\lambda_{j}^{ ~\dot a}=-
\tilde\lambda_i^{~\dot a} \tilde\lambda_{j ~\dot a}.$ In conventions of
\cite{Dixon}, equation \eqref{scprod} would have a minus sign on the right hand side.

An on-shell momentum
of a massless particle, $p_{k}^\mu$ can be written as
\be
p_{k~a \dot a} =\ p_{k\mu}\, \sigma^\mu =\ \lambda_{k~a}\, \tilde\lambda_{k~\dot a} \ .
\ee
In section 3 we use a Lorentz-invariant
combination
$\vev{ i^-|{{p\!\!\!/}\;}_k|j^-} =i^a \,p_{k~a\dot a}\,j^{\dot a}$,
which in terms of the spinor products \eqref{Aspicon} is
\be \label{lorinv}
\vev{ i^-|{{p\!\!\!/}\;}_k|j^-} = \
\vev{ i^-|^a \ |k^+}_a \ \vev{k^+|_{\dot a} \ |j^-}^{\dot a} = \
-\vev{i~k} \, [k~j] = \ \vev{i~k} \, [j~k] \ .
\ee

The spinors $\lambda$ and $\tilde{\lambda}$ appearing in the
helicity formalism are precisely the wave-functions
of fermions of positive and negative helicities,
\SP{
\label{spwavefs}
\langle i^-|^{a}= \ \lambda_i^{~a}=\ \overline{u_-}(k_i)^a \ , \quad
&|i^+\rangle_a =\ \lambda_{i~a}=\ u_{+}(k_i)_a \ ,
\\
\langle i^+|^{\dot a}= \ \tilde\lambda_i^{~\dot a}=\ \overline{u_+}(k_i)^{\dot a}
 \ , \quad
&|i^-\rangle_{\dot a} =\ \tilde\lambda_{i~\dot a}=\ u_{-}(k_i)_{\dot a} \ .
}
In our conventions for MHV vertices we treat all fermions
and antifermions as incoming and the fermion propagator connects two
incoming fermions with opposite helicities. Thus, the completeness relation
relevant for us and with a correct index structure gives
\be \label{comprelcor}
{|i^+\rangle_a  \, |i^-\rangle_{\dot a} \over k_i^2} =\
{\lambda_{i~a} \, \tilde\lambda_{i~\dot a} \over  k_i^2} =\
{{{k\!\!\!/}}_{i~ a \dot a} \over k_i^2} \ ,
\ee
which is, of course, the correct fermion propagator in usual
Feynman perturbation theory.
%The abbreviation $|-i\rangle$ stands for
%the spinor corresponding to an incoming momentum $-k_i$.

Scalars have no wave-functions, and their propagator remains
$1/k^2,$ and vectors give (in Feynman gauge)
\be
{\varepsilon_{+}^\mu  \, \varepsilon_{-}^\nu \over k^2}=\
{-g^{\mu\nu} \over k^2} \ ,
\ee
which is the correct form of the massless vector boson propagator.

An important consequence of \eqref{comprelcor} is the ordering prescription
of fermions in MHV vertices. This concerns  only the
case of scalar diagrams with  internal fermion lines, such as the third
diagram in Figure 1.
In this case, in order to get the
ket$^+$ ket$^-$ structure $|i^+\rangle_a |-i^-\rangle_{\dot a}$
the two fermions which are to be connected by a propagator
should be both on the right of each vertex (rather than adjacent to each other).
This means that, for example, the third diagram in Figure 1 comes from
\be \label{Apthreeef}
A(\Lambda_1^-,g_2^-,\underline{\Lambda_{(3i)}^+})
\ A(g_3^-,\Lambda_k^+,\underline{\Lambda_{-(3i)}^-})
\ .
\ee
If the contracted fermion factors, $\underline{\Lambda_{(3i)}^+}$ and
$\underline{\Lambda_{-(3i)}^-}$
were, instead, chosen to be next to each other, the overall contribution would change
sign, since fermions anticommute with each other.

\bigskip
%%%%%%%%%%%%%%%%%%%%%%%%%%%%%%%%%%%%%%%%%%%%%%%%%%%%

%%%%%%%%%%%%%%%%%%%%%%%%%%%%%%%%%%%%%%%%%%%%%%%%%%

%%%%%%%%%%%%%%%%%%%%%%%%%%%%%%%%%%%%%%%%%%%%%%%%%%%%%%%%%%%
%\newpage

\end{document}